\documentclass[preprint,12pt]{elsarticle}

\usepackage{amsmath}
\usepackage{bm}
\usepackage[a4paper, top=2.5cm, bottom=2.8cm, right=2.4cm, left=2.4cm]{geometry}

\usepackage{amssymb,color}
\usepackage{hyperref}

\newcommand{\bra}[1]{\langle #1 |}
\newcommand{\ket}[1]{| #1 \rangle}
\newcommand{\bk}[1]{\langle  #1 \rangle}

\newcommand{\tr}{{\rm tr}}
\newcommand{\e}{{\rm e}}
\newcommand{\Z}{{\mathcal{Z}}}

\newcommand{\s}{{\rm scal}}
\newcommand{\sq}{{\rm spin}}

\newcommand{\Gc}{{\mathcal{G}}}
\newcommand{\Gcd}{{\dot{\mathcal{G}}}}

\newcommand{\half}{\frac{1}{2}}

\begin{document}

\begin{frontmatter}



\title{Low-energy limit of N-photon amplitudes\\ in a constant field: Part II}


\author{Misha A. Lopez-Lopez}

   \address{Helmholtz-Zentrum Dresden-Rossendorf, Bautzner Landstraße 400, 01328 Dresden, Germany}

   \address{Institut für Theoretische Physik, Technische Universität Dresden, 01062 Dresden, Germany}            
             

\begin{abstract} 

We employ the worldline formalism to derive a series representation of the low-energy limit of the $N$-photon amplitude in a constant background field for both scalar and spinor QED.
The amplitudes are then written in terms of a single proper-time integral. 
The above-mentioned series representation terminates when considering a constant crossed field. This allows us to obtain even more compact expressions for these particular amplitudes for which the result of the proper-time integral, for fixed parameters, takes the form of a factorial function. In addition, we derive all helicity components of these amplitudes and  express them explicitly in terms of Bernoulli numbers and spinor products. 

\end{abstract}


\end{frontmatter}


\section{Introduction}
\label{intro}
Photon amplitudes have been an active subject of study since the early estimations of the photon-photon scattering cross section \cite{euler-kockel,euler,akhiezer} and the seminal papers by Euler-Heisenberg \cite{eulhei} and Weisskopf \cite{weisskopf}. 
Currently, in vacuum, these amplitudes have been computed at one-loop level for up to six photons with arbitrary photon energies  \cite{karneu50,karneu51,detollis-64,detollis-65,costantinitopi,Ahmadiniaz4photon,bernicotgui}, and for any number of photons in the low-energy limit \cite{Martin,Dunne}. 
At the two-loop level, assuming every photon has low energy, the amplitudes have been computed for up to ten photons with arbitrary helicities and for an arbitrary  number of photons when all helicities are equal  \cite{Martin,Dunne,Edwards-low}.
In addition, there are a few nonperturbative results of photon amplitudes in the presence of background fields for two \cite{Narozny,Baier76,Mauren,DiPiazzaMerging,DiPiazzaMerging2}, three \cite{adler,adler17,Papanyan,DiPiazzaSplit,DiPiazzaDelbruck} and four \cite{Baier2018,Ahmadiniaz-inprep} photons.

The worldline formalism \cite{Schubert-rep,Corradini-wl} has also made possible the derivation of compact expressions for the $N$-photon amplitudes in the presence of a constant field in the low-energy photon regime \cite{Ahmadiniaz-Nphotonlow} (referred as ``Part I"). 

It is important to mention that the worldline formalism has been employed to obtain, for both scalar and spinor QED, master formulas for the one-loop $N$-photon amplitudes in vacuum \cite{polyakov-book,Bern-Kosower,strassler1,Schubert-rep,Ahmadiniaz4photon}. 
These were later generalized to include the nonperturbative contributions of a constant  \cite{Shaisultanov-NphC,Reuter}, plane-wave \cite{Edwards-NphPW}, and  combined constant  and plane-wave  \cite{Schubert-Shaisultanov} background fields 
(for an alternative derivation of these amplitudes in constant and plane-wave fields, see Refs. \cite{Baier75,DiPiazza:2013iwa}). 
Although these master formulas are valid for off-shell photons of arbitrary energy and any field strengths, the exact contribution to physical processes can be extracted only after calculating $N$ proper-time integrals which in most cases are non-trivial.

In Part I (see also Refs. \cite{Dunne,Edwards-low}), we showed that in the low-energy limit of photons, it is possible to considerably simplify these $N$-photon amplitudes to the point in which only a single proper-time integral is left. 
In this work, we continue studying the $N$-photon amplitudes in a constant background field \cite{Schubert-rep,Shaisultanov-NphC,Reuter} for low-energy photons. We show that in the presence of a constant crossed field, the $N$-photon amplitudes involve only a single proper-time integral which, for fixed parameters, takes the form of a factorial function. Furthermore, by following Ref. \cite{Martin}, we can completely determine the helicity components of these amplitudes in terms of spinor products.

We work in the Euclidean $(g^{\mu\nu}) = {\rm diag}(+1,+1,+1,+1)$, however, for the calculation of the helicity components, we employ the conventions of Refs. \cite{Martin,Dixon-sh} where the metric tensor in Minkowski space is $(\eta^{\mu\nu}) = {\rm diag}(+1,-1,-1,-1)$.

\section{Low-energy limit of the $N$ - photon amplitudes in vacuum}

In this section, we  summarize the derivation of the helicity components for the low-energy limit of the on-shell $N$-photon amplitudes in vacuum, adapted from Ref. \cite{Martin}.  This derivation makes use of the one-loop effective Lagrangians in a constant background field \cite{eulhei,weisskopf} and  the spinor helicity formalism \cite{Dixon-sh} (see also Ref. \cite{Ahmadiniaz-hel}).  This is in preparation for the following sections, where similar results are obtained in the presence of a constant crossed field.

In order to obtain the polarized $N$-photon amplitudes it is necessary to perform the following steps, see Ref. \cite{Martin} for details. First, consider  the  Euler-Heisenberg Lagrangian for spinor QED \cite{eulhei}\footnote{From now on, we use natural units with $c=\hbar=1$, $m$ represents the mass of the particle in the loop  and we use the absolute value of the electron charge $e = |e|$.}
\begin{equation}\label{EH}
	\mathcal{L}^{}_{\rm EH}=-\frac{1}{8\pi^2}\int_0^\infty \frac{dT}{T^3}{\rm e}^{-m^2T} \left\{\frac{(e a T)(e b T)}{\tan(e a T)\tanh ({\it ebT})} -\frac{2}{3}(eT)^2 \mathcal{F}-1 \right\} \,,
\end{equation}
and the Weisskopf Lagrangian for scalar QED \cite{weisskopf} 
\begin{equation}\label{weisskopf}
	\mathcal{L}^{}_{\rm W}=\frac{1}{16\pi^2}\int_0^\infty \frac{dT}{T^3}{\rm e}^{-m^2T} \left\{\frac{(e a T)(e b T)}{\sin(e a T)\sinh ({\it ebT})} +\frac{1}{6}(eT)^2 \mathcal{F}-1 \right\}\,,
\end{equation}
where
\begin{equation}\label{max-inv}
	a=\left(\sqrt{\mathcal{F}^2+\mathcal{G}^2}-\mathcal{F} \right)^{1/2}~,\hspace{1.5cm} b=\left(\sqrt{\mathcal{F}^2+ \mathcal{G}^2}+\mathcal{F}\right)^{1/2},
\end{equation}
with the two invariants of the Maxwell field
\begin{equation}
	-2\mathcal{F}=-\frac{1}{2}F_{\mu\nu}F^{\mu\nu}= \vec{E}^2-\vec{B}^2\,,\hspace{1.5cm}
	-\mathcal{G}=-\frac{1}{4}F_{\mu\nu}\tilde{F}^{\mu\nu} =\vec{E}\cdot\vec{B}\, . 
\end{equation}
Second, fix the field strength tensor as 
\begin{equation}\label{Ftoti}
	F^{\mu\nu}  = f_{\rm tot}^{\mu\nu} = \sum_{i=1}^N f_i^{\mu \nu} \,,
\end{equation}
where 
\begin{equation}\label{fi}
	f_i^{\mu \nu} = k_i^\mu \varepsilon_i^\nu - \varepsilon_i^\mu k_i^\nu
\end{equation} 
represents the field strength tensor of an external photon  with four-momentum $k_i^\mu$ and polarization four-vector $\varepsilon_i^\mu$. 
Then, the corresponding $N$-photon amplitude, for low-energy (LE) photons, 
is obtained by extracting the terms involving each $f_1,..., f_N$ precisely once
\begin{equation}\label{amplitude-Ftot}
	\begin{split}
		\Gamma_{\left\{{\sq\atop \s}\right\}}^{({\rm LE})} 	(k_1,\varepsilon_1;\ldots ;k_N,\varepsilon_N)= 
		\mathcal{L}_{\left\{{EH\atop W}\right\}}(if_{\rm tot})\Big|_{f_1\cdots f_N} \,.
	\end{split}		
\end{equation}
The low-energy limit of the photon amplitudes is defined by the condition that all photon energies are much smaller than the mass of the particles in the loop, therefore requiring all kinematic invariants $k_i\cdot k_j$ to satisfy $k_i\cdot k_j\ll m^2$.

Third, note that the spinor representation of the polarization vector for an on-shell photon of definite helicity $\pm1$ is 
\begin{equation}
	\varepsilon_\mu^{\pm}(k) = \pm \, \frac{\bra{q^\mp}\gamma_\mu \ket{k^\mp}}{\sqrt{2}\, \langle q^\mp|k^\pm \rangle} \,,
\end{equation}
where $\langle q^\pm|k^\mp \rangle = \overline{u_{\pm}(q)}\,u_{\mp}(k)$  are spinor products, and $q$ is a reference momentum. The spinor products are represented in the usual form: 
\begin{equation}
	\bk{ij} = \langle k_i^-|k_j^+ \rangle = \overline{u_{-}(k_i)}\,u_{+}(k_j) \,,\qquad
	[ij] = \langle k_i^+|k_j^- \rangle = \overline{u_{+}(k_i)}\,u_{-}(k_j)  \,.
\end{equation}

Fourth, assume fixed polarization for the photons, with $L$ having the helicity `$+$' and $N-L$ the helicity `$-$', and note that the two Maxwell invariants can be expressed as
\begin{equation}
	\frac{1}{4} f_{{\rm tot}\,\mu\nu} f_{\rm tot}^{\mu\nu} = \chi_+ + \chi_- \,,\qquad
	\frac{1}{4} f_{{\rm tot}\,\mu\nu} \tilde{f}_{\rm tot}^{\mu\nu} = -i(\chi_+ - \chi_- )\,,
\end{equation}
where
\begin{equation}
	\chi_+ = \half\, \sum_{1\leq i<j\leq N}\, [ij]^2  \,,\qquad
	\chi_- = \half\, \sum_{1\leq i<j\leq N}\, \bk{ij}^2 \,,
\end{equation}
such that the invariants in \eqref{max-inv} become
\begin{equation}
	a=-i(\sqrt{\chi_+} - \sqrt{\chi_-} ) \,, \qquad b= \sqrt{\chi_+} + \sqrt{\chi_-} \;.  
\end{equation}

Fifth, insert these expressions into the Euler-Heisenberg Lagrangian \eqref{EH} and expand the trigonometric functions as Taylor series to obtain
\begin{equation}\label{LEHspin}
	\begin{split}
		\mathcal{L}_{\rm EH}(if_{\rm tot})  = -2 \,\frac{m^4}{(4\pi)^{2}}\sum_{N=4}^\infty \left(\frac{2e}{m^2}\right)^N 
		\sum_{L=0\atop L\,{\rm even}}^N  \, c_{\sq}^{(1)}(N,L)\; \chi^{\frac{L}{2}}_+ \; \chi^{\frac{N-L}{2}}_- \,,
	\end{split}		
\end{equation}
where
\begin{equation}\label{Cspinb}
	\begin{split}
		c_{\sq}^{(1)}(N,L)\, = (-1)^{N/2} \,(N-3)!\, \sum_{r=0}^L \sum_{s=0}^{N-L}\,(-1)^{N-L-s}\, \frac{{\cal B}_{r+s} \,{\cal B}_{N-r-s}}{r!\,s!\,(L-r)!\,(N-L-s)!}
	\end{split}		
\end{equation}
and ${\cal B}_{n}$ are Bernoulli numbers. Similarly, the Weisskopf Lagrangian  \eqref{weisskopf} can be expressed as
\begin{equation}\label{LWscal}
	\begin{split}
		\mathcal{L}^{}_{\rm W}(if_{\rm tot})  = \frac{m^4}{(4\pi)^{2}}\sum_{N=4}^\infty \left(\frac{2e}{m^2}\right)^N 
		\sum_{L=0 \atop L\,{\rm even}}^N  \, c_{\s}^{(1)}(N,L)\; \chi^{\frac{L}{2}}_+ \; \chi^{\frac{N-L}{2}}_- \,,
	\end{split}		
\end{equation}
where   
\begin{equation}\label{Cscalb}
	\begin{split}
		c_{\s}^{(1)}(N,L)\, =  (-1)^{N/2} \,(N-3)!\, \sum_{r=0}^L \sum_{s=0}^{N-L}\,(-1)^{N-L-s}\, \frac{\Big(1-2^{1-r-s}\Big)\Big(1-2^{1-N+r+s}\Big) \,{\cal B}_{r+s} \,{\cal B}_{N-r-s}}{r!\,s!\,(L-r)!\,(N-L-s)!}\,.
	\end{split}		
\end{equation}

Finally, according to \eqref{amplitude-Ftot}, the amplitudes with $L$ `$+$' and $N - L$ `$-$' helicities are obtained from the corresponding term in the sum of \eqref{LEHspin} and \eqref{LWscal} by picking out the terms multilinear in the $f_i$'s. For $L$ even, it is defined \cite{Martin}
\begin{equation}\label{chi-plus}
	\begin{split}
		\chi_{L}^+ \, = (\chi_+)^{\frac{L}{2}} |_{\textrm{all different}} = \frac{\left(L/2\right)!}{2^{L/2}} \Big\{[12]^2[34]^2\cdots [(L-1)L]^2 + \textrm{all permutations}\Big\}\,,
	\end{split}		
\end{equation}
\begin{equation}\label{chi-minus}
	\begin{split}
		\chi_{N-L}^- \, &= (\chi_-)^{\frac{N-L}{2}} |_{\textrm{all different}}\\
		& = \frac{\left(\frac{N-L}{2}\right)!}{2^{\frac{N-L}{2}}} \Big\{\bk{(L+1)(L+2)}^2\bk{(L+3)(L+4)}^2\cdots \bk{(N-1)N}^2 + \textrm{all permutations}\Big\}\,.
	\end{split}		
\end{equation}
Therefore, the low-energy limit of the $N$-photon amplitudes with $L$ external photons having helicity `$+$' and $N-L$ `$-$' are
\begin{equation}\label{hspin-vacuum}
	\begin{split}
		\Gamma_{\left\{{\s\atop \sq}\right\}}^{(LE)}(f_{1}^+;...;f_{L}^+ f_{L+1}^-;...;f_{N}^-)  = \left\{{1\atop -2}\right\} \,\frac{m^4}{(4\pi)^{2}} \left(\frac{2e}{m^2}\right)^N \, c_{\left\{{\s\atop \sq}\right\}}^{(1)}(N,L)\; \chi_{L}^+ \; \chi_{N-L}^- \;.
	\end{split}		
\end{equation}

In general, the low-energy limit of an $N$-photon amplitude is given by the first non-vanishing term of its inverse-mass expansion.
In Ref. \cite{Martin}, it has been noted that the leading contributions  \eqref{hspin-vacuum} to the $N$-photon amplitudes obey a ``double Furry theorem". 
First, the charge symmetry of the theory (Furry theorem) restricts the number of external photons $N$ to be even.
Second,  if there is an odd number of positive or negative helicities, these  contributions vanish.

\section{Low-energy limit of the $N$ - photon amplitudes in a constant field}\label{CFsec}
In this section, we consider the one-loop off-shell $N$-photon amplitudes in a constant field for photon energies much smaller than the electron mass. Here, we follow the conventions of Ref. \cite{Schubert-rep} and part I.

In part I, we have shown that the low-energy limit of the $N$-photon amplitude in a constant background field for scalar QED  can be expressed as 
\begin{equation}\label{N-phot20}
	\begin{split}
		\Gamma_{\s}^{(LE)} (k_1,\varepsilon_{1};... ;k_N,\varepsilon_{N};F)  & =  \frac{e^N}{(4\pi)^{\frac{D}{2}}}   
		\int_{0}^{\infty} \frac{dT}{T} \, T^{N-\frac{D}{2}} \e^{-m^2 T} \, {\rm det}^{1/2}\left[\frac{\Z}{\sin \Z}\right]\\
		&\quad\times\exp\left\{ \sum_{n=1}^{\infty} \frac{1}{2n}\,I_{\s}^{\rm cyc}(f_{\rm tot},\dots,f_{\rm tot};F)\right\}\Bigg|_{f_1...f_N}\,,
	\end{split}		
\end{equation}
where $\Z^{\mu\nu} = eTF^{\mu\nu}$, turning the problem of computing the amplitude into the calculation of the cyclic integral
\begin{equation}\label{cyclicint}
	\begin{split} 
		I_\s^{\rm cyc}(f_1,f_2,\dots,f_n;F) &= \int_0^1 du_1 \cdots \int_0^1 du_n\, \Gcd_{B}(12\dots n) \,,
	\end{split}		
\end{equation}
with the bosonic Lorentz trace:
\begin{equation}\label{Blonrentz}
	\Gcd_{B}(i_1i_2\dots i_n) = \left(\frac{1}{2}\right)^{\delta_{n1}+\delta_{n2}} \tr (f_{i_1}\cdot \Gcd_{Bi_1i_2}\cdot f_{i_2}\cdot \Gcd_{Bi_2i_3}\cdots f_{i_n}\cdot \Gcd_{Bi_ni_1} )\,.
\end{equation}
Here, the first derivative of the `bosonic' worldline Green's function in the presence of a constant field $\Gcd_{Bij}$ can be represented  as a Neumann series expansion in terms of the differential operator \cite{Schubert-rep,Hassani} as follows:\footnote{The subscript `$P$' stands for periodic boundary conditions.}
\begin{equation}\label{caldotG}
	\begin{split}
		&\Gcd_{B}(u_i,u_j) = 2 \bra{u_i} \left(\partial_{P} - 2i\Z \right)^{-1}\ket{u_j} 
		= 2\sum_{\ell = 0}^{\infty} \bra{u_i} \partial_{P}^{-(\ell+1)} \ket{u_j} (2i\Z)^\ell\,.
	\end{split}
\end{equation}
For such representation, we can define a generic bosonic cycle integral, as in the vacuum case \cite{Ahmadiniaz4photon}, by 
\begin{equation}\label{binte}
	\begin{split} 
		b_{\ell_{1}+...+\ell_{n}} = 2^{\ell_{1}+...+\ell_{n}}\,\int_0^1 du_1\, du_2 \cdots du_n\, \bra{u_1} \partial_{P}^{-\ell_1} \ket{u_2} \bra{u_2} \partial_{P}^{-\ell_2} \ket{u_3}
		\cdots \bra{u_n} \partial_{P}^{-\ell_n} \ket{u_1}   \,.
	\end{split}		
\end{equation}
The calculation of the previous integral follows from the completeness relation $\int_0^1 du\, \ket{u}\bra{u} = 1$ and it can be expressed in terms of the Bernoulli numbers ${\cal B}_\ell$ \cite{Edwards:2021elz,Schmidt}
\begin{equation}\label{Bernoullin}
	b_\ell = \quad\left\{ 
	\begin{array}{ll}
		-2^\ell \,{{\cal B}_\ell\over \ell!}  & \qquad \ell{\rm \quad even}\,,\\
		0 & \qquad \ell{\rm \quad odd}\,.\\
	\end{array}\right.
\end{equation}
This allows us to obtain a series expansion representation for the cyclic integral
\begin{equation}\label{Iscal}
	\begin{split} 
		I_\s^{\rm cyc}(f_1,f_2,\dots,f_n;F) & = \left(\frac{1}{2}\right)^{\delta_{n1} + \delta_{n2}}\, \sum_{\ell_1 = 0}^{\infty} \,\cdots \,\sum_{\ell_n = 0}^{\infty} i^{\ell_{1}+...+\ell_{n}}\,
		b_{n+\ell_{1}+...+\ell_{n}} \\
		&\quad \times \tr \left(f_1 \cdot \Z^{\ell_1} \cdot f_2 \cdot \Z^{\ell_2} \cdots f_n \cdot \Z^{\ell_n} \right) \,.
	\end{split}		
\end{equation}

\subsection{Spinor QED}

According to the ``replacement rule" (see Ref. \cite{Schubert-rep} and part I), the low-energy limit of the $N$-photon amplitude in a constant field for spinor QED is
\begin{equation}
	\begin{split}
		\Gamma_{\sq}^{(LE)}(k_1,\varepsilon_{1};...;k_N,\varepsilon_{N};F) & = -2 \frac{e^N}{(4\pi)^{\frac{D}{2}}}   
		\int_{0}^{\infty} \frac{dT}{T} \, T^{N-\frac{D}{2}} \e^{-m^2 T} \, {\rm det}^{1/2}\left[\frac{\Z}{\tan \Z}\right]\\
		&\quad \times
		\exp\left\{ \sum_{n=1}^{\infty} \frac{1}{2n}\,I_\sq^{\rm cyc}(f_{\rm tot},\dots,f_{\rm tot};F)\right\}\Bigg|_{f_1...f_N}\,,
	\end{split}		
\end{equation}
where now the spinor cyclic integral is
\begin{equation}
	\begin{split} 
		I_\sq^{\rm cyc}(f_1,f_2,\dots,f_n;F) &= \int_0^1 du_1 \cdots \int_0^1 du_n\, \Big[\Gcd_{B}(12\dots n) - \Gc_{F}(12\dots n) \Big]\,,
	\end{split}		
\end{equation} 
which contains both the bosonic Lorentz trace \eqref{Blonrentz} and its fermionic  counterpart
\begin{equation}\label{Florentz}
	\Gc_{F}(i_1i_2\dots i_n) = \left(\frac{1}{2}\right)^{\delta_{n1}+\delta_{n2}} \tr (f_{i_1}\cdot \Gc_{Fi_1i_2}\cdot f_{i_2}\cdot \Gc_{Fi_2i_3}\cdots f_{i_n}\cdot \Gc_{Fi_ni_1} )\,,
\end{equation}
where the fermionic worldline Green's function in the presence of a constant field $\Gc_{Fij}$ can be expressed as\footnote{The subscript `$A$' stands for anti-periodic boundary conditions.}
\begin{equation}\label{calF}
	\begin{split}
		&\Gc_{F}(u_i,u_j) = 2 \bra{u_i} \left(\partial_{A} - 2i\Z \right)^{-1}\ket{u_j} 
		= 2\sum_{\ell = 0}^{\infty} \bra{u_i} \partial_{A}^{-(\ell+1)} \ket{u_j} (2i\Z)^\ell\,.
	\end{split}
\end{equation}
Similar to the scalar case, see Eq. (\ref{binte}), the fermionic cycle integral can be expressed in terms of Bernoulli numbers by \cite{Edwards:2021elz,Schmidt}
\begin{equation}
	\begin{split} 
		2^{\ell_{1}+...+\ell_{n}}\,\int_0^1 du_1 \cdots \int_0^1 du_n\, \bra{u_1} \partial_{A}^{-\ell_1} \ket{u_2}  
		\cdots \bra{u_n} \partial_{A}^{-\ell_n} \ket{u_1} = \left(1-2^{\ell_{1}+...+\ell_{n}}\right) b_{\ell_{1}+...+\ell_{n}}\,.
	\end{split}		
\end{equation}
Then, in this case, the spinor cyclic integral becomes
\begin{equation}\label{Ispin}
	\begin{split} 
		I_\sq^{\rm cyc}(f_1,f_2,\dots,f_n;F) & = \left(\frac{1}{2}\right)^{\delta_{n1} + \delta_{n2}}\, \sum_{\ell_1 = 0}^{\infty} \,\cdots \,\sum_{\ell_n = 0}^{\infty}i^{\ell_{1}+...+\ell_{n}}\, h_{n+\ell_{1}+...+\ell_{n}}
		 \\
		&\quad \times  \,\tr \left(f_1 \cdot \Z^{\ell_1} \cdot f_2 \cdot \Z^{\ell_2} \cdots f_n \cdot \Z^{\ell_n} \right) \, 
	\end{split}		
\end{equation} 
where we have set $h_\ell = (2-2^\ell) b_\ell $.

\section{Low-energy limit of the $N$ - photon amplitudes in a constant crossed field}\label{CCFsec}
In this section, we consider a constant crossed field, in four dimensions, defined by $\bm{E} \perp \bm{B}$, $|\bm{E}| = |\bm{B}|$.  Since the electric and magnetic fields are perpendicular and equal in magnitude both invariants  $\bm{E} \cdot \bm{B}$ and $|\bm{B}|^2 - |\bm{E}|^2$ vanish. This implies $F^3 = 0$  so that the power series in (\ref{Iscal}) and (\ref{Ispin})  terminate after their quadratic term.  
Then, the determinants become
\begin{equation}\label{Det-ccf}
	{\rm det}^{1/2}\left[\frac{\Z}{\tan \Z}\right] =  
	{\rm det}^{1/2}\left[\frac{\Z}{\sin \Z}\right] = 1\,.
\end{equation}

The Lorentz traces, in this case, can be further simplified by noticing that  $F^2 \cdot f_i \cdot F^2 = 0$, which restricts the number of $F$'s that can appear in a Lorentz trace to be less than or equal to the number of $f_i$'s  in such trace.
Then, the cyclic integrals \eqref{Iscal} and \eqref{Ispin} can be represented as
\begin{equation}\label{Is-sq}
	\begin{split} 
		I_{\left\{\s\atop \sq\right\} }^{\rm cyc}(f_1,f_2,\dots,f_n;F) & = \left(\frac{1}{2}\right)^{\delta_{n1} + \delta_{n2}}\, 
		\sum_{\ell = 0}^{n} \,(ieT)^{\ell}\, \left\{{b_{n+\ell}\atop h_{n+\ell}}\right\} \,\\
		&\quad\times \tr^{\rm dist} \left( f_1 \cdot F \cdot f_2\cdot F  \cdots f_\ell\cdot F \cdot f_{\ell+1}\cdot f_{\ell+2} \cdots f_n\right) \,,\\
	\end{split}		
\end{equation}
where `$\tr^{\rm dist} $' denotes the sum of all inequivalent permutations of $F$ for a fix set of $f_i$'s. 
For instance, the non-null contributions, for $n=1$
\begin{equation}
	\tr^{\rm dist}(f_1 \cdot F) = \tr(f_1 \cdot F) \,.
\end{equation}
For $n= 2$
\begin{equation}
	\begin{split}  
		\tr^{\rm dist} \left(f_1 \cdot F \cdot f_2\cdot F\right)  &= 
		\tr( F \cdot f_1 \cdot F\cdot f_2) + \tr(  F^2 \cdot f_1 \cdot f_2) + \tr( f_1 \cdot  F^2 \cdot f_2) \,.
	\end{split}		
\end{equation}
And for $n=3$
\begin{equation}
	\begin{split}  
		&\tr^{\rm dist} \left( f_1\cdot F  \cdot f_2 \cdot f_3\right)  = 
		\tr( f_1 \cdot  F \cdot f_2 \cdot f_3) + \tr( f_1  \cdot f_2\cdot  F \cdot f_3)+ \tr(f_1 \cdot f_2 \cdot f_3\cdot  F) \,,\\
		&\tr^{\rm dist} \left( f_1 \cdot F  \cdot f_2\cdot F  \cdot f_3\cdot F \right)  = \left(f_1 F f_2 F f_3 F\right)  + \tr \left(f_1 f_2 F f_3 F^2 + 5 \,{\rm perm ~ of~} F,\, F^2 \right)\,.
	\end{split}		
\end{equation}

Then, after replacing the value of the determinants, the low-energy limit of the $N$-photon amplitudes in a constant crossed field are 
\begin{equation}\label{Nlow-ccf}
	\begin{split}
		\Gamma_{\left\{{\s\atop \sq}\right\}}^{(LE)} (k_1,\varepsilon_{1};... ;k_N,\varepsilon_{N};F)  & = \left\{{1\atop -2}\right\}  \frac{e^N}{(4\pi)^{2}}   
		\int_{0}^{\infty} \frac{dT}{T} \, T^{N-2} \e^{-m^2 T} \\
		&\quad\times \exp\left\{ \sum_{n=1}^{\infty} \frac{1}{2n}\,
		I_{\left\{\s\atop \sq\right\}}^{\rm cyc}(f_{\rm tot},\dots,f_{\rm tot};F)\right\}\Bigg|_{f_1...f_N}\,.
	\end{split}		
\end{equation}
It is important to note that these $N$-photon amplitudes are valid off-shell and the proper-time integral left is straightforward to perform for any fixed number of photons.

\subsection{Helicity components}\label{sec-hely-ct}
Given that, a constant crossed field is the low-frequency approximation of a plane-wave field, its vector-potential can be expressed as $A(\phi) = E_0(\varepsilon_0^{+} + \varepsilon_0^{-}) \phi$ where $\phi = n_\mu x^\mu$, $n^\mu = (1,\bm{n})$, $\bm{n}$ is a unitary vector defining the propagation direction of the plane wave, $E_0$ is a constant equal in magnitude to the electric and magnetic field strengths and $\varepsilon_0^{\pm,\mu}$ are unitary four vectors orthogonal to $n^\mu$ that, in this case, we identify as the `$+$' and `$-$' helicity components of the plane wave field \cite{DiPiazzaSplit,Barducci-ccf}. Then, the field strength tensor of a constant crossed field is $F = f_0^+ + f_0^- $ where $f_{0}^{\pm,\mu\nu} = k_0^\mu \,\varepsilon_0^{\pm,\nu} - \varepsilon_0^{\pm,\mu}k_0^\nu $ and $k_0^\mu = E_0 \,n^\mu$.

On the other hand, in Refs. \cite{Martin,Ahmadiniaz-hel}, it is found that the  following commutator and anticommutator relations hold for on-shell photons with fixed helicity
\begin{equation}
	[f_i^{+},f_j^{-}]^{\mu\nu}= 0 \,,
\end{equation}
\begin{equation}
	\{f_i^{+},f_j^{+}\}^{\mu\nu} = - \frac{1}{2} [ij]^2\, \eta^{\mu\nu}\,,
\end{equation}
\begin{equation}
	\{f_i^{-},f_j^{-}\}^{\mu\nu} = - \frac{1}{2} \bk{ij}^2\, \eta^{\mu\nu}\,,
\end{equation}
and that these relations imply the factorization of traces, such that 
\begin{equation}\label{factor-taces}
	\tr(f_{i_1}^{+}\cdots f_{i_M}^{+} f_{j_1}^{-} \cdots f_{j_N}^{-}) = \frac{1}{4}\tr(f_{i_1}^{+}\cdots f_{i_M}^{+})\,\tr(f_{j_1}^{-} \cdots f_{j_N}^{-}) \,,
\end{equation}
where the same-helicity traces are
\begin{equation}\label{traces-pol}
	\tr(f_{i_1}^{+}\cdots f_{i_N}^{+}) = \frac{(-1)^N}{\sqrt{2^{N-2}}}\, 
	[i_1i_2][i_2i_3]\, \cdots \,[i_{N}i_{1}] \,,
\end{equation}
\begin{equation}\label{traces-pol2}
	\tr(f_{i_1}^{-}\cdots f_{i_N}^{-}) = \frac{(-1)^N}{\sqrt{2^{N-2}}}\, \bk{i_1i_2}\bk{i_2i_3}\, \cdots \,\bk{i_{N}i_{1}}  \,.
\end{equation}

These relations restrict the contributions of $f_0^\pm$ in the $N$-photon amplitudes in a constant crossed field, when $L$ external photons have helicity `$+$' and $N-L$ `$-$', there are at most $L$ contributions of $f_0^+$ and $N-L$ of $f_0^-$. Then, by summing all possible combinations of $f_0^\pm$, the helicity components of Eq. \eqref{Nlow-ccf} can be expressed as
\begin{equation}\label{expansion-low}
	\begin{split}
		&\Gamma_{\left\{{\s\atop \sq}\right\}}^{(LE)}(f_1^+;...;f_{L}^+ ;f_{L+1}^{-}; ...;f_{N}^-;F)  = \left\{{1\atop -2}\right\}\frac{(-i)^N}{(4\pi)^{2}}  \, 
		\int_{0}^{\infty} \frac{dT}{T^3} \, \e^{-m^2 T} \, 
		\sum_{n=0}^{N-L} \sum_{\ell=0}^L\\
		&\quad\times \exp\left\{ \sum_{r=1}^{\infty}   \left(\frac{1}{2}\right)^{\delta_{r2}}\, \frac{(ieT)^{r}}{2r} 
			\left\{{b_{r}\atop h_{r}}\right\}
			\,\tr\left[ \left( f_{\rm tot}^{\prime} \right)^r\right] \right\} \Bigg|_{f_1^+...f_{L}^+  f_{L+1}^{-} ...f_{N}^-f_{N+1}^{+} ... f_{N+\ell}^{+}f_{N+\ell+1}^{-}...f_{N+\ell +n}^{-}}\,,
	\end{split}		
\end{equation}
where we have introduced 
\begin{equation}\label{ftotpi}
	(f_{\rm tot}^{\prime})^{\mu\nu} = \sum_{i=1}^{2N} f_i^{\mu \nu} \,
\end{equation}
to take into account the $N$ copies of $f_0$ by fixing $f_{N+1} = f_{N+2} = ... =f_{2N} = f_0$. 

Comparing the results in Refs. \cite{Martin,Dunne}, we can see that the second line of \eqref{expansion-low} is equal to\footnote{Specifically, compare Eq. \eqref{hspin-vacuum} with Eqs. (18) and (20) of Part I.}
\begin{equation}\label{ct-factor}
	(2ieT)^{N+n+\ell}\,c_{\left\{{\s\atop \sq}\right\}}^{(1)}(N+n+\ell,L+\ell) \;\frac{ \chi_{L+\ell}^+ \; \chi_{N-L+n}^-}{(N+n+\ell -3)!}\,,	
\end{equation}
where $c_{\s}^{(1)}(N,L)$, $c_{\sq}^{(1)}(N,L)$ and $\chi_{M}^\pm$ were given in \eqref{Cscalb}, \eqref{Cspinb}, \eqref{chi-plus} and \eqref{chi-minus}, respectively.

Therefore, the low-energy limit of the $N$-photon amplitudes in a constant crossed field with $L$ external photons having helicity `$+$' and $N-L$ `$-$' are
\begin{equation}\label{hscal}
	\begin{split}
		\Gamma_{\left\{{\s\atop \sq}\right\}}^{(LE)}(f_{1}^+;...;f_{L}^+ f_{L+1}^-;...;f_{N}^-;F)  &= {\left\{{1\atop -2}\right\}} \frac{m^4}{(4\pi)^{2}} \left(\frac{2e}{m^2}\right)^N \;
		\sum_{n=0}^{N-L} \sum_{\ell=0}^{L} \left(\frac{2ie}{m^2}\right)^{n+\ell} \\ 
		& \quad\times c_{\left\{{\s\atop \sq}\right\}}^{(1)}(N+n+\ell,L+\ell) \; \chi_{L+\ell}^+ \; \chi_{N-L+n}^- \,.
	\end{split}		
\end{equation}
In these polarized amplitudes, we can observe that the 'double Furry theorem' is determined not by the external photons but by every contribution to the amplitudes. 
This is, of course, due to the presence of the constant crossed field, which introduces the sum over $\ell$ and $n$ contributions of  $f_0^+$ and  $f_0^-$, respectively.
Then, it is convenient to think of $\chi_{L+\ell}^+$ as the sum of inequivalent spinor products of $L$  external photons with '$+$'  helicity and $\ell$ field contributions $f_0^+$ (similarly for $\chi_{N-L+n}^-$). 
For instance, let us consider the following two cases: 

\begin{enumerate}
	\item[(i)] For $N=3$, $L=2$, the only non-vanishing contributions are $\chi_{2}^+ \, \chi_{1+1}^- = \frac{1}{4}\, [12]^2 \bk{30}^2$ and $\chi_{2+2}^+ \, \chi_{1+1}^- = \frac{1}{4}\, [10]^2 [20]^2 \bk{30}^2$:
	\begin{equation}
		\begin{split}
			\Gamma_{\left\{{\s\atop \sq}\right\}}^{(LE)}(f_{1}^+;f_{2}^+;f_{3}^-;F)  &= {\left\{{1\atop -2}\right\}} \frac{m^4}{(4\pi)^{2}} \left(\frac{2e}{m^2}\right)^4 \; 
			\left(\frac{i}{4}\right)\Bigg[c_{\{ \cdot \}}^{(1)}(4,2)\, [12]^2 \bk{30}^2 \\
			&\quad - \left(\frac{2e}{m^2}\right)^2 c_{\{ \cdot \}}^{(1)}(6,4)\, [10]^2 [20]^2 \bk{30}^2 \Bigg]\,.
		\end{split}		
	\end{equation}

	\item[(ii)] For $N=4$, $L=2$, the non-vanishing contributions are all possible combinations of  $\chi_{2}^+  = \frac{1}{2}\, [12]^2 $ and  $\chi_{2+2}^+  = \frac{1}{2}\, [10]^2 [20]^2 $   with  $\chi_{2}^-  = \frac{1}{2}\, \bk{34}^2$ and $\chi_{2+2}^-  = \frac{1}{2}\, \bk{30}^2\bk{40}^2$:
	\begin{equation}
		\begin{split}
			\Gamma_{\left\{{\s\atop \sq}\right\}}^{(LE)}(f_{1}^+;f_{2}^+;f_{3}^-;&f_{4}^-;F)  = {\left\{{1\atop -2}\right\}} \frac{m^4}{(4\pi)^{2}} \left(\frac{2e}{m^2}\right)^4 \;
			\left(\frac{1}{4}\right)\Bigg[c_{\{ \cdot \}}^{(1)}(4,2)\, [12]^2 \bk{34}^2 \\
			&\quad - \left(\frac{2e}{m^2}\right)^2 \left(c_{\{ \cdot \}}^{(1)}(6,4)\, [10]^2 [20]^2 \bk{34}^2 + c_{\{ \cdot \}}^{(1)}(6,2)\, [12]^2 \bk{30}^2 \bk{40}^2\right)\\
			&\quad+ \left(\frac{2e}{m^2}\right)^4 c_{\{ \cdot \}}^{(1)}(8,4)\, [10]^2 [20]^2 \bk{30}^2 \bk{40}^2\Bigg]\,.
		\end{split}		
	\end{equation}
\end{enumerate}

\section{Summary and Outlook}

To summarize, within the worldline approach, we have extended our study of the low-energy limit of the $N$-photon amplitudes in an arbitrary constant field for both scalar and spinor QED. 
We have represented these amplitudes as multiple series expansions with respect to the background field such that only a single proper-time integral is left.
By fixing the external field as a constant crossed field, we have obtained compact explicit expressions for these amplitudes as well as for their helicity components. 

\section*{Acknowledgments}
I am grateful to  C. Schubert for helpful comments and suggestions on the present work. I thank N. Ahmadiniaz, A. Di Piazza and  R. Schützhold, for a careful and critical manuscript reading.

\bigskip

%




\end{document}